\documentclass[oldversion]{aa}
\usepackage{epsfig}
\usepackage{natbib}
\usepackage{lscape}
\usepackage{wasysym}
\bibpunct{(}{)}{;}{a}{}{,}
\begin{document}

\title{The HARPS search for southern extra-solar planets\thanks{Based on 
observations collected at the La Silla Parana Observatory,
ESO (Chile) with the HARPS spectrograph at the 3.6m telescope, under
the GTO program 072.C-0488.}}

\subtitle{XII. A giant planet orbiting the metal-poor star HD\,171028}

\author{
  N. C. Santos\inst{1,2} \and
  M. Mayor\inst{2} \and
  F. Bouchy\inst{3} \and
  F. Pepe\inst{2} \and
  D. Queloz\inst{2} \and
  S. Udry\inst{2}
}

\institute{
    Centro de Astrof{\'\i}sica, Universidade do Porto, Rua das Estrelas, 
    P--4150-762 Porto, Portugal
    \and
    Observatoire de Gen\`eve, 51 ch. des Maillettes, CH--1290 Sauverny, Switzerland
    \and
Institut d'Astrophysique de Paris, UMR7095 CNRS, Université Pierre \& Marie Curie, 98bis Bd Arago, 75014 Paris, France 
}

\date{Received 2007-06-21; accepted 2007-08-07}

\abstract{
In this paper we present the detection of a 1.8\,M$_{\mathrm{Jup}}$ planet 
in a 538\,day period trajectory orbiting the metal-poor 
star \object{HD\,171028} ([Fe/H]=$-$0.49). This planet is the first to be 
discovered in the context of a HARPS program searching for planets around 
metal-poor stars. Interestingly, HD\,171028 is one of the least metal-poor stars 
in the sample. This discovery is placed in the context of the models of planet 
formation and evolution.
  \keywords{stars: individual: HD\,171028 -- 
            stars: planetary systems --
	    planetary systems: formation -- 
            techniques: radial-velocity -- 
	    stars: abundances
	    }}

\authorrunning{Santos et al.}
\maketitle

\section{Introduction}

The discovery of more than 220 extra-solar planets orbiting
solar-type stars\footnote{See e.g. table at http://www.exoplanets.eu/} is providing
crucial evidence for the processes of planet formation and evolution 
\citep[for a review see e.g.][]{Udry-2007}.
Among these, important clues come from the study of planet-host stars.
The well known strong correlation between the presence of giant planets and the 
stellar metallicity \citep[][]{Gonzalez-1997,Gonzalez-2001,Santos-2001,Santos-2004b,Santos-2005a,
Reid-2002,Fischer-2005}
suggests that giant planets are more easily formed around higher metal-content stars.
This lends supports to the core-accretion model \citep[e.g. ][]{Mizuno-1980,Pollack-1996} as the main planet formation mechanism \citep[e.g.][]{Ida-2004b,Benz-2006}, in opposition to the alternative disk-instability 
model \citep[][]{Boss-1997,Boss-2002,Mayer-2002}.

The higher probability of finding giant planets orbiting metal-rich stars
prompted a number of different surveys dedicated to metal-rich samples \citep[e.g.][]{Tinney-2002,Fischer-2005b,DaSilva-2006,Melo-2007}. Given their
observing strategy, these programs unveiled mostly short period planets,
strongly biasing the known samples, while positively increasing the number
of detected transiting planets orbiting bright stars \citep[e.g.][]{Sato-2005,Bouchy-2005b}.

Interestingly, however, several giant planets were found to orbit metal-poor stars,
with down to $\sim$2 times less metals than the Sun \citep[e.g.][]{Setiawan-2003,Mayor-2004,Cochran-2007}. Some programs to search 
for planets around such objects were also started. Two of these include the use of
both Keck and HET telescopes \citep[][respectively]{Sozzetti-2006b,Cochran-2007}. A
third one, presented in this paper, is part of the HARPS GTO planet search
program \citep[][]{Mayor-2003b}.

Part of the goal of these programs is to try to understand how frequent are giant
planets orbiting metal-poor stars, and what is the metallicity limit below which no giant
planets {can be observed}. Such constraints would give important new clues about the processes
of planet formation and evolution \citep[e.g.][]{Matsuo-2007}. The recent finding that the metallicity correlation may no longer be valid for Neptune-mass planets 
(see case of HD\,4308 ([Fe/H]=$-$0.31) and discussion in \citet[][]{Udry-2006}),
together with theoretical predictions suggesting that very low mass planets 
may be common around metal-poor stars \citep[][]{Ida-2004a,Benz-2006}, renewed 
the interest for these surveys.

In this paper we present the first detection of a giant planet orbiting one star from
the HARPS metal-poor stars survey. The planet orbits \object{HD\,171028}
([Fe/H]=$-$0.49) in a 538\,day period orbit. In Sect.\,\ref{sec:sample} we describe the sample
used. In Sect.\ref{sec:planet} we present the observations of \object{HD\,171028}, 
providing the stellar parameters and the orbital solution found. We conclude 
in Sect.\,\ref{sec:discussion}.

\section{The HARPS metal-poor sample}
\label{sec:sample}

The HARPS GTO program started in October 2003 to follow several different samples
of solar-type stars \citep[][]{Mayor-2003b}. The remarkable long term precision
of HARPS allowed the discovery of several planets among the targets
\citep[e.g.][]{Pepe-2004,Lovis-2005}, including the large majority
of the known planets with masses of the order of the mass of Neptune or 
below \citep[][]{Santos-2004a,Bonfils-2005b,Bonfils-2007,Lovis-2006,Udry-2006,Udry-2007b}.

To explore the low metallicity tail of the planet-host stars distribution, 
one of the samples currently studied with HARPS is constituted of 105 
metal-poor or mild metal-poor solar-type stars. This sample was chosen based on the large FGK-catalogue of \citet[][]{Nordstrom-2004}. From this catalogue, we took all
late-F, G, and K stars ($b-y>$0.330) south of $+$10$^{\mathrm{o}}$ of declination and 
having a visual V magnitude brighter than 12. From these we then excluded all known visual
and spectroscopic binaries, all stars suspected to be giants, and all those 
with measured projected rotational velocity $v\,\sin{i}$ above $\sim$6.0\,km\,s$^{-1}$ 
(to indirectly exclude the most active stars). Finally, we considered 
only those targets with photometric [Fe/H] between $-$0.5 and $-1.5$. 

\begin{figure}[t]
\resizebox{\hsize}{!}{\includegraphics{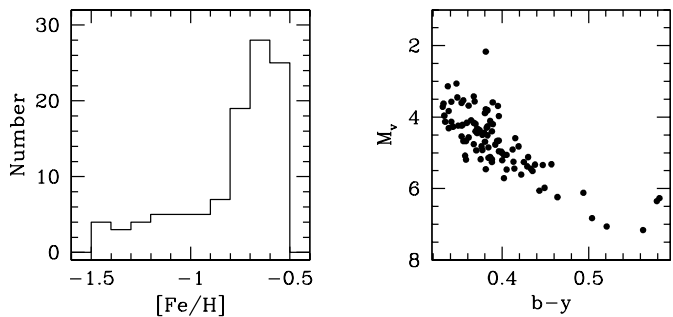}}
\caption{{\it Left}: Metallicity distribution for the whole HARPS metal-poor 
sample {(photometrically derived metallicities)}. {\it Right}: M$_v$ vs. $b-y$ diagram for the stars in the sample.}
\label{fig:sample}
\end{figure}

In Fig.\,\ref{fig:sample} we plot the metallicity distribution of the sample, 
as well as a M$_v$ vs. $b-y$ diagram. The distribution is clearly peeked 
between $-$0.5 and $-$0.8, while a long ``flat'' metallicity tail is present 
down to [Fe/H]=$-$1.5. We note that HD\,171028 (spectroscopic metallicity of 
$-$0.49), was taken into the catalogue only because its photometric 
metallicity is lower than this value (see discussion in the next section).

The final 105 stars in the sample have their V magnitudes between 5.9 and 10.9, 
distributed around an average value of 8.7. After a 15 minutes exposure
these magnitudes allow to obtain a S/N high enough to derive radial-velocities
with a precision better than 1\,m\,s$^{-1}$ for the majority of the targets.

\section{A new planet around HD\,171028}
\label{sec:planet}

\subsection{Observations}
\label{sec:observations}

We obtained a total of 19 measurements of HD\,171028 in GTO time (program ID 072.C-0488) using 
the HARPS spectrograph (3.6-m ESO telescope, La Silla, Chile). The observations
were carried out between October 2004 and
April 2007, and the radial-velocities were obtained using the latest version
of the HARPS pipeline. We refer to \cite{Pepe-2004} for details on the data reduction. 

Exposure times varied between 260 and 900\,s, and the individual photon-noise error 
in the radial-velocities was always below 1.7\,m\,s$^{-1}$ (median of 0.8\,m\,s$^{-1}$).
In this error we do not quantify the uncertainties due to the stellar oscillation
modes \citep[][]{Bouchy-2005c}. This ``noise'' is typically averaged out if a long
(15 minutes) exposure is done. Unfortunately, this was not always the case
when observing HD\,171028 (only 12 out of the 19 measurements were done following
this strategy).

The individual spectra were also used to derive both the Bisector Inverse Slope (BIS) of the 
HARPS Cross-Correlation Function (CCF), as defined by \cite{Queloz-2000}, 
as well as a measurement of the chromospheric activity index $\log{R'_{\rm HK}}$, 
following a similar recipe as used by \citet[][]{Santos-2000a} for
CORALIE spectra. Finally, the combined high S/N HARPS spectra were analyzed to
derive stellar atmospheric parameters and iron abundances using the method described 
in \citet[][]{Santos-2004b}.

\subsection{Stellar characteristics}

There is little information available on HD\,171028 (BD$+$06\,3833, V$=$8.31, B-V$=$0.61), 
and only two papers appear in a Simbad query \citep[][]{Olsen-1994,Nordstrom-2004}.
The star is also not in the Hipparcos catalogue {\citep[][]{ESA-1997}}, and most of the information
available on the Tycho catalogue is not very accurate (e.g. the parallax listed
is 9.1$\pm$7.8\,milliarcsec).

\begin{table}
\par
\caption{
\label{table:hd171028_star}
Stellar parameters for \object{HD\,171028}. }
\begin{tabular}{lcc}
\hline\hline
\noalign{\smallskip}
Parameter  & Value & Reference \\
\hline
Spectral~type   		& G0 			& Simbad  \\
$m_v$           		& 8.31 			& Simbad \\
$B-V$  				& 0.61                  & Simbad \\
$b-y$  				& 0.43                  & \citet[][]{Nordstrom-2004} \\
$\log{R'_{\rm HK}}$ 		& $-$4.92$\dagger$      & This paper \\
Distance~[pc]   		& 90                    & This paper \\
$v~\sin{i}$~[km~s$^{-1}$] 	& 2.3$\dagger\dagger$ 	& This paper \\
$T_{\rm eff}$~[K]  		& 5663$~\pm~$20 	& This paper \\
$\log{g}$  			& 3.84$~\pm~$0.05 	& This paper \\
$\xi_{\mathrm{t}}$  		& 1.32$~\pm~$0.05 	& This paper \\
${\rm [Fe/H]}$  		& $-$0.49$~\pm~$0.02 	& This paper \\
Mass~$[M_{\odot}]$  		& 0.99$\pm$0.08         & This paper \\
\hline
\noalign{\smallskip}
\end{tabular}
\newline
$\dagger$ From HARPS spectra using a calibration similar to the one presented 
by \citet{Santos-2000a};
\newline
$\dagger\dagger$ From HARPS spectra using a calibration similar to the one presented 
by \citet{Santos-2002a}
\end{table}

The analysis of our combined HARPS spectra, with a total S/N ratio above 500, 
provide $T_{eff} = 5663\pm20$, $\log g=3.84\pm0.05$, and $[Fe/H]=-0.49\pm0.02$. 
These values were obtained following the methodology and line-lists used
in \citet[][]{Santos-2004b}. Similar values for the stellar parameters
are obtained using the larger \ion{Fe}{i} and \ion{Fe}{ii}
line-list presented in \citet[][]{Sousa-2007}, and the automatic
ARES\footnote{http://www.astro.up.pt/$\sim$sousasag/ares} code for line-equivalent width measurement
($T_{eff} = 5693\pm16$, $\log g=3.85\pm0.05$, and $[Fe/H]=-0.48\pm0.01$).
For the rest of the paper we will consider the first set of parameters.
{The errors mentioned above denote internal errors only. Systematic errors
affecting e.g. the temperature scale are thoroughly discussed in the literature \citep[e.g.][]{Santos-2004b,Ramirez-2004,Casagrande-2006,Masana-2006}, and no 
consensus seems to exist at this
point. This discussion is, however, out of the scope of the current paper.}

The effective temperature and surface gravity derived are compatible with 
the expected parameters of a slightly evolved solar-type star, being roughly 
compatible with the spectral type of G0 listed in the Simbad database.
Using the method described in \citet[][]{Pont-2004b} with the isochrones
of \citet[][]{Girardi-2000} we derived a stellar mass and age of
0.99$\pm$0.08\,M$_\odot$ and 6-11 Gyr, respectively. The stellar
radius inferred is 1.95$\pm$0.26\,R$_\odot$. {The
uncertainties in these parameters were
derived using 2-sigma errors in the effective temperature, surface gravity
and metallicity.}
Using this mass and radius together with the derived surface gravity and 
effective temperature, and making use of the bolometric correction 
of \citet[][]{Flower-1996}, we derive a distance of 90\,pc to HD\,171028, 
with an error around 15\,pc if we consider a conservative error
in the $\log g$ of 0.15\,dex. No reddening corrections were taken into
account in this estimate.

The values mentioned above are a bit different from those listed in the
\citet[][]{Nordstrom-2004} catalogue ($T_{eff} = 5432\,K$, $[Fe/H]=-0.81\pm0.02$,
Mass=0.78\,M$_{\odot}$, and distance of 43\,pc). We note, however, that the photometric calibrations used by Nordstr\"om et al. to derive these parameters (including the absolute
magnitude used to derive the distance) are likely only valid for main-sequence
stars.

From the HARPS spectra we derived a chromospheric activity index ($\log{R'_{\rm HK}}=-4.92$, with rms of 0.02) following a procedure similar to the one presented in \citep[][]{Santos-2000a}.
From the activity level and the $B-V$ colour we 
estimate a rotational period of 19\,days \citep[][]{Noyes-1984} and
an age of 4\,Gyr \citep[][]{Henry-1996} \citep[or at least above 2\,Gyr --][]{Pace-2004}. 
{These latter values are likely not accurate due to the fact that HD\,171028 
is a metal-poor star and a bit evolved off the main sequence \citep[][]{Wright-2004}.
The calibrations mentioned above were based on main-sequence stars, mostly of solar metallicity.
In any case, the measured chromospheric activity level is compatible with HD\,171028 being an old, 
chromospherically quiet star.
}

\begin{table}
\caption{Chemical abundances for the several alpha and iron peek elements
studied. For titanium, both results based on \ion{Ti}{i} and \ion{Ti}{ii} 
lines are presented. The abundances given by \ion{Fe}{i} and \ion{Fe}{ii}
lines is the same \citep[it is one of the conditions for the derivation of
the stellar parameters --][]{Santos-2004b}. The error in [X/H] represents the rms of the
abundances given by the n(X) different lines used.}
\label{table:elements}
\begin{tabular}{lcccc}
\hline\hline
Element & log$\epsilon_\odot$ & [X/H] & n(X) & [X/Fe] \\
\hline
Fe   &  7.47 & $-$0.49$\pm$0.02 & 38 & 0.00\\
Si   &  7.55 & $-$0.43$\pm$0.02 & 11 & 0.06\\
Ca   &  6.36 & $-$0.41$\pm$0.05 & 13 & 0.08\\
Sc   &  3.10 & $-$0.40$\pm$0.05 & 6  & 0.09\\
\ion{Ti}{i}  &  4.99 & $-$0.44$\pm$0.03 & 12 & 0.05\\
\ion{Ti}{ii} &  4.99 & $-$0.43$\pm$0.04 & 4  & 0.06\\
V    &  4.00 & $-$0.54$\pm$0.04 & 8  & $-$0.05\\
Cr   &  5.67 & $-$0.56$\pm$0.02 & 4  & $-$0.07\\
Mn   &  5.39 & $-$0.72$\pm$0.06 & 4  & $-$0.23\\
Co   &  4.92 & $-$0.50$\pm$0.03 & 6  & $-$0.01\\
Ni   &  6.25 & $-$0.53$\pm$0.03 & 29 & $-$0.04\\
Na   &  6.33 & $-$0.42$\pm$0.07 & 3  & 0.07\\
Mg   &  7.58 & $-$0.44$\pm$0.03 & 2  & 0.05\\
Al   &  6.47 & $-$0.46$\pm$0.01 & 2  & 0.03\\
\hline
\end{tabular}
\end{table}

\subsubsection{Abundances of different species}

It is known that thick-disk stars have typically higher abundances
of alpha elements \citep[][]{Bensby-2003,Fuhrmann-2004}. 
To try to access if HD\,171028 could be a member of this galactic
population, we used the method and line-lists described in \citet[][]{Gilli-2006} and \citet[][]{Santos-2006b} to derive the abundances of several alpha and
iron-peek elements for HD\,171028. 
The results, listed in Table\,\ref{table:elements}, suggest that this star
has typical abundances of thin disk solar-type stars \citep[see also][]{Gilli-2006}.
In particular, no overabundance of alpha elements is seen when comparing with 
stars of similar [Fe/H].

No trace of the Li line at 6707.8\,\AA\ is found in our S/N$\sim$500
spectrum of HD\,171028. 
{An upper limit of 0.6\,m\AA\ was obtained for the Equivalent Width of
the Li-line. This value translates into an upper limit of $\log{\epsilon}(Li)<0.2$\,dex. 
Such a value is compatible with its evolutionary 
status and effective temperature, although a large dispersion is seen in the Li
abundances of sub-giant stars similar to {HD\,171028} \citep[][]{Randich-1999,Lebre-1999}. 
}

\subsection{HARPS orbital solution}

As mentioned above, we obtained 19 accurate radial-velocity measurements of \object{HD\,171028}
with the HARPS spectrograph, between October 2004 and April 2007. The complete radial velocity measurements obtained and the corresponding errors are presented in Table\,\ref{table:rv}. It is worth noticing that the errors quoted in the table, which are used to plot the error bars, refer solely to the instrumental (calibration) and photon-noise error 
share of the total error budget (e.g. activity and/or stellar oscillations are not considered, 
given the difficulty in having a clear estimate of their influence).

\begin{table}[b]
\caption{HARPS radial-velocity measurements of HD\,171028.}
\label{table:rv}
\begin{tabular}{ccc}
\hline\hline
JD     &$V_r$ [km\,s$^{-1}$] &$\sigma(V_r)$ [km\,s$^{-1}$]\\
\hline
2453310.4946  &  13.6319  &  0.0009\\
2453574.6582  &  13.5948  &  0.0008\\
2453575.6376  &  13.5924  &  0.0007\\
2453576.6401  &  13.5899  &  0.0008\\
2453577.6084  &  13.5939  &  0.0008\\
2453578.6591  &  13.5954  &  0.0007\\
2453669.5027  &  13.7099  &  0.0016\\
2453672.4942  &  13.7045  &  0.0013\\
2453862.8107  &  13.6268  &  0.0011\\
2453864.8024  &  13.6304  &  0.0011\\
2453870.7806  &  13.6268  &  0.0009\\
2453882.8334  &  13.6266  &  0.0010\\
2453883.8141  &  13.6219  &  0.0017\\
2453920.7281  &  13.6179  &  0.0008\\
2454166.9019  &  13.6329  &  0.0009\\
2454168.8885  &  13.6345  &  0.0008\\
2454172.8740  &  13.6417  &  0.0008\\
2454174.8916  &  13.6446  &  0.0008\\
2454196.8946  &  13.7075  &  0.0007\\
\hline
\end{tabular}
\end{table}

\begin{figure}[t!]
\resizebox{\hsize}{!}{\includegraphics{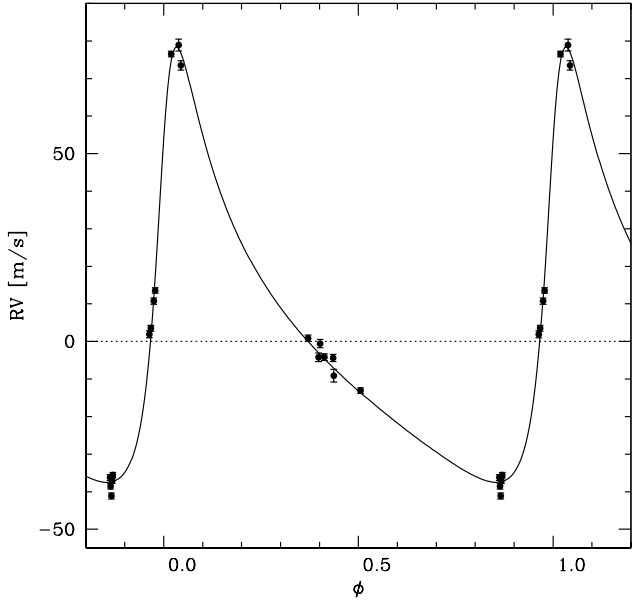}}
\caption{{\it Top}: Phase-folded radial-velocity measurements of HD\,171028, and
the best Keplerian fit to the data with a period of 538-days, eccentricity of 0.61,
and semi-amplitude of 58\,m\,s$^{-1}$. }
\label{fig:plot_phase}
\end{figure}

\begin{figure}[t]
\resizebox{\hsize}{!}{\includegraphics{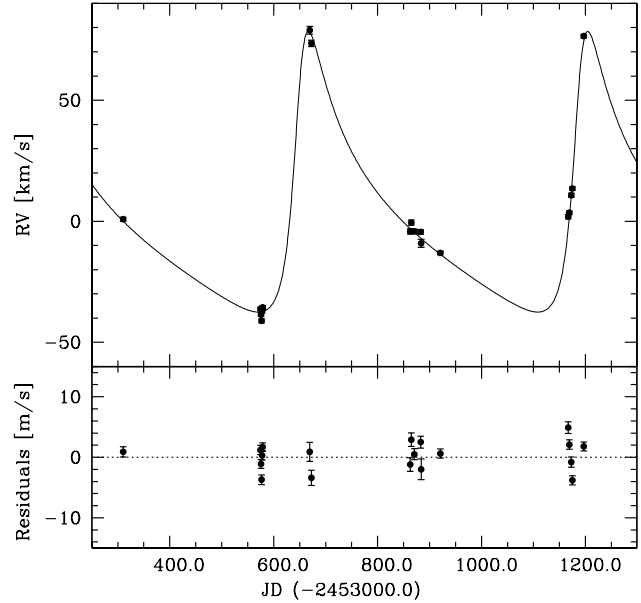}}
\caption{{\it Top}: Radial-velocity measurements of HD\,171028 as a function of time, and
the best Keplerian fit to the data. {\it Bottom}: Residuals of the fit.}
\label{fig:plot_time}
\end{figure}

Just after the first measurements were done the star was noticed to be
radial-velocity variable. A later analysis of the whole data set revealed the 
presence of a 538-day period radial-velocity signal. This signal is 
best fitted using a Keplerian fit with an semi-amplitude K of 58\,m\,s$^{-1}$, 
and an eccentricity of 0.61 (see Figs.\,\ref{fig:plot_phase} and \ref{fig:plot_time}). 
Given the mass for HD\,171028, this corresponds to the expected signal 
induced by the presence of a 1.83 Jupiter-masses (minimum-mass) 
companion (Table\,\ref{table:hd171028_orbit}).

\begin{table}[b]
\caption[]{Elements of the fitted orbit for HD\,171028b.}
\begin{tabular}{lll}
\hline
\hline
\noalign{\smallskip}
$P$             & 538$\pm$2				& [d]\\
$T$             & 2453648.9205$\pm$1.8584		& [d]\\
$a$		& 1.29					& [AU]\\
$e$             & 0.61$\pm$0.01				&  \\
$V_r$           & 13.631$\pm$0.001			& [km\,s$^{-1}$]\\
$\omega$        & 305$\pm$1				& [degr] \\ 
$K_1$           & 58.0$\pm$0.4				& [m\,s$^{-1}$] \\
$f_1(m)$        & 5.421\,10$^{-9}$			& [M$_{\odot}$]\\ 
$\sigma(O-C)$   & 2.4					& [m\,s$^{-1}$]  \\    
$N$             & 19					&  \\
$m_2\,\sin{i}$  & 1.83					& [M$_{\mathrm{Jup}}$]\\
\noalign{\smallskip}
\hline
\end{tabular}
\label{table:hd171028_orbit}
\end{table}

\begin{figure}[t!]
\resizebox{\hsize}{!}{\includegraphics{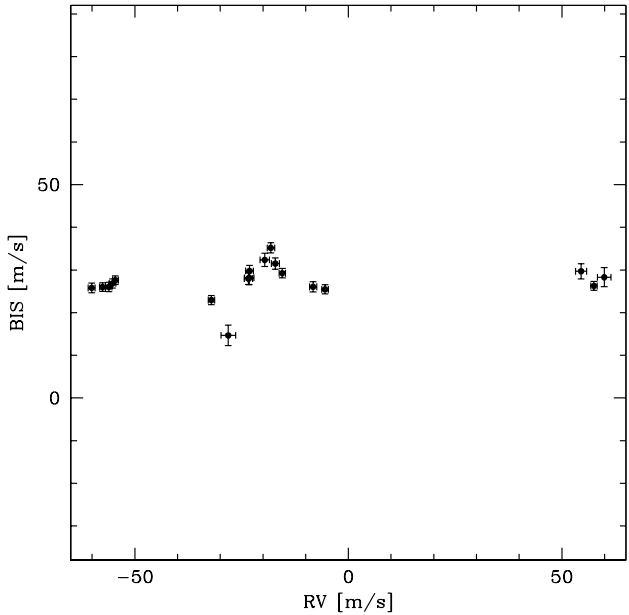}}
\caption{BIS vs. radial-velocity for HD\,171028. To strongly evidence the nonexistence of a 
correlation the vertical and horizontal scales were set to be the same.}
\label{fig:vrbis}
\end{figure}

To understand if the periodic radial-velocity signal observed could have a non-planetary 
origin \citep[see e.g.][]{Saar-1997,Queloz-2000,Santos-2002a}, in Fig.\,\ref{fig:vrbis} 
we plot the Bisector Inverse Slope (BIS) of the Cross-Correlation Function 
as a function of the radial-velocity. The result shows that
no correlation exist between the two quantities, suggesting that stellar activity
of stellar blends cannot explain the radial-velocity variation observed.
Together with the low activity level of the star, we conclude that the
538-day orbital period observed can be better explained by the presence of
a Jupiter-like planet orbiting HD\,171028.

The residuals of the orbital fit have a rms of 2.4\,m\,s$^{-1}$, slightly above the
average photon-noise error of the measurements (0.95\,m\,s$^{-1}$). 
The lower panel of the plot presented in Fig.\,\ref{fig:plot_time} does reveal 
some structure in the residuals after the 538-day period orbit is
subtracted. Although we cannot exclude that this signal may be
due to the presence of another companion to the system, the fact that 
some structure is also present in the plot of Fig.\,\ref{fig:vrbis} may
hint at a non-planetary origin for the signal. More data are needed
to settle this issue.

\section{Concluding remarks}
\label{sec:discussion}

In this paper we present the detection of the first planet among the stars
in the HARPS metal-poor planet search program, a giant planet with a minimum
mass of 1.8\,M$_{\mathrm{Jup}}$ orbiting HD\,171028 in an eccentric trajectory
every 538 days.

This detection adds to the small number of planets known to orbit stars
with metallicity clearly below solar \citep[e.g.][]{Mayor-2004,Cochran-2007}. 
While the correlation between the presence of planets and stellar metallicity 
is clearly established \citep[e.g.][]{Gonzalez-2001,Santos-2001,Santos-2004b,Fischer-2005}, the 
detection of an increasing number of giant planets orbiting low-metallicity
stars reopens the debate about the origin of these worlds. These findings 
show that giant planet formation is not completely inhibited 
around stars in the intermediate metal-poor regime. 

These cases either represent the metal-poor tail of the giant planets formed by the
core-accretion process, or they may hint at the existence of a
different population of planets formed as a result of disk-instability processes. 
The precise study of the metallicity distribution of stars with planets 
suggests that there may exist a ``long'' flat low-metallicity 
tail \citep[][]{Santos-2004b,Udry-2007}. Although the statistics of
metal-poor planet-hosts is still poor, we can speculate that 
we may be observing a superposition of two populations. On the one hand,
stars whose planets were formed by the metallicity dependent core-accretion 
process \citep[][]{Pollack-1996,Ida-2004b}, mostly populating the metal-rich 
regime. On the other hand, a less significant population of giant planets
orbiting stars of all metallicities. These latter could have been formed
by the disk-instability process \citep[][]{Boss-2002}. We are considering here
that subsequent planet evolution processes (e.g. migration in the disk)
are not strongly dependent on the metallicity \citep[][]{Livio-2003}. 

We should add, however, that according to the models, giant planets formed
by the disk-instability process could have higher masses when compared
with those formed by core-accretion \citep[e.g.][]{Matsuo-2007}. The 
lack of a clear correlation between planet-mass and stellar 
metallicity \citep[e.g.][]{Santos-2003} may be an important caveat 
for the proposed scenario.

In this context it is interesting to verify that HD\,171028 is clearly
one of the most metal-rich stars in the HARPS sample presented in this paper.
Although statistically not relevant 
{(note that our sample is clearly more populated at
the metallicity range between $-$0.5 and $-$0.8), if confirmed this fact could lend some 
support to the core-accretion model. 
}

{
Also interesting is the fact that HD\,171028 is a bit evolved off the main-sequence.
In a very recent study, \citet[][]{Pasquini-2007} present evidence that giant
stars with planets are likely not as metal-rich as their dwarf counterparts \citep[see also ][]{LicioDaSilva-2006}.
Although other explanations exist \citep[e.g. the higher mass of the stars, and eventually of the 
proto-planetary disks, may significantly change the planet formation efficiency --][]{Laughlin-2004,Endl-2006,Bonfils-2007,Johnson-2007}, 
this interesting conclusion 
could suggest that planetary pollution may be more important than previously
thought \citep[e.g.][]{Pinsonneault-2001,Santos-2003,Santos-2004b,Fischer-2005}. 
We note, however, that no clear connection seems to be present between the stellar 
evolution status and [Fe/H] among dwarfs and sub-giant stars
with planets \citep[e.g.][]{Santos-2003,Fischer-2005}. In the sub-giant branch, 
planet-host stars are still metal-rich 
when compared with ``single'' field stars, even though important dilution processes
may have occurred. A look at the tables of \citet[][]{Santos-2004b,Santos-2005a} 
and \citet[][]{Sousa-2006}, also shows that low metallicity 
among planet hosts is not restricted to evolved stars. Such an observations could 
be expected if stellar pollution was a frequent outcome of the 
planet formation process. Out of 14 stars with metallicity below $-$0.20\,dex, 
only 5 are likely evolved ($\log{g}<$4.2),
while 7 ``definite'' dwarf stars ($\log{g}>$4.4) are also metal-poor. 
On the other hand, several evolved stars exist with metallicity well above solar.
Finally, although some caveats have been discussed \citep[][]{Vauclair-2004}, the 
lack of any correlation between stellar
metallicity and convective envelope mass is still an important argument against
pollution being the main mechanism responsible for the observed metal rich nature
of planet-host stars \citep[e.g.][]{Pinsonneault-2001,Santos-2003,Fischer-2005}. 
The K-dwarf stars with planets, with deep convective regions, would require an excessive infall of 
planetary material to get enriched to the observed level.
}

The continuation of the program presented here will certainly provide 
important constraints for this discussion. On the one hand, it will help
to understand what is the lower stellar metallicity limit for which giant planets
are still able to form \citep[see discussion in][]{Matsuo-2007}. 
On the other hand, with the adequate observing
strattegy, it will allow to access the frequency of Neptune-mass planets
orbiting lower metallicity stars \citep[see][]{Udry-2006}. Although
the (unknown) disk masses may also have a crucial influence on the planet formation
efficiency, such results would have clearly important implications for the models 
of planet formation and evolution.

\begin{acknowledgements} 
We would like to thank F. Pont for the help in deriving the stellar mass, age
and radius, as well as to our referee L. Pasquini, who helped to improve
the clarity of the paper. This work made use of the Simbad database. 
NCS would like to thank the support from Funda\c{c}\~ao para a Ci\^encia 
e a Tecnologia, Portugal, in the form of a grant (reference POCI/CTE-AST/56453/2004). This work 
was supported in part by the EC's FP6 and by FCT (with
POCI2010 and FEDER funds), within the HELAS international
collaboration.
\end{acknowledgements}

\bibliographystyle{aa}
\bibliography{santos_bibliography}

\end{document}